\documentclass[fleqn,usenatbib]{mnras}

\usepackage{newtxtext,newtxmath}
\usepackage[T1]{fontenc}
\usepackage{ae,aecompl}

\usepackage{graphicx}	% Including figure files
\usepackage{dblfloatfix}
\usepackage{amsmath}	% Advanced maths commands
\usepackage{threeparttable}
\usepackage{subcaption}
\usepackage{caption}
\usepackage{color}
\usepackage[dvipsnames]{xcolor}
\usepackage[normalem]{ulem}
\usepackage{booktabs}

\newcommand{\fermi}{\textit{Fermi}-{\rm LAT}}
\newcommand{\planck}{\textit{Planck}}

\newcommand{\gray}{$\gamma$-ray}
\newcommand{\grays}{$\gamma$-rays}

\newcommand{\xrays}{$\rm X$-rays}
\newcommand{\etacarina}{$\eta$ Car}

\newcommand{\msun}{\mbox{$M_\odot$}}

\def\deg{\hbox{$^\circ$}}

\title[\fermi\ detection around the Carina Nebula Complex ]{Diffuse \gray\ emission around the massive star forming region of Carina Nebula Complex}
\author[Ge et.al]{Ting-Ting Ge$^{1}$, 
Xiao-Na Sun$^{1}$\thanks{E-mail:xiaonasun@gxu.edu.cn}, 
Rui-Zhi Yang$^{2}$$^{3}$$^{4}$, 
Yun-Feng Liang$^{1}$,
En-Wei Liang$^{1}$
\\
$^{1}$Guangxi Key Laboratory for Relativistic Astrophysics, School of Physical Science and Technology, Guangxi University, Nanning 530004, China\\
$^{2}$Department of Astronomy, School of Physical Sciences, University of Science and Technology of China, Hefei, Anhui 230026, China\\
$^{3}$CAS Key Labrotory for Research in Galaxies and Cosmology, University of Science and Technology of China, Hefei, Anhui 230026, China\\
$^{4}$School of Astronomy and Space Science, University of Science and Technology of China, Hefei, Anhui 230026, China\\
}

\pubyear{2022}

\begin{document}
\label{firstpage}
\pagerange{\pageref{firstpage}--\pageref{lastpage}}
\maketitle

% Abstract of the paper
\begin{abstract}
We report the Fermi Large Area Telescope (\fermi) detection of the \gray\ emission toward the massive star forming region of Carina Nebula Complex (CNC).
Using the latest source catalog and diffuse background models, we found that the GeV \gray\ emission in this region can be resolved into three different components.
%morphological and spectral analyses allow us to resolve the \gray\ emission into three.
%
%One of the GeV \gray\ emission from suggest a colliding-wind massive star binary systems $\eta$ Carina ($\eta$ Car).
The GeV \gray\ emission from the central point source is considered to originate from the $\eta$ Carina ($\eta$ Car). % as previous suggested \citep{Mart21}. 
We further found the diffuse GeV \gray\ emission around the CNC which can be modelled by two Gaussian disks with radii of 0.4$\deg$ (region A) and 0.75$\deg$ (region B), respectively. 
The GeV \gray\ emission from both the regions A and B have good spatial consistency with the derived molecular gas in projection on the sky.
The GeV \gray\ emission of region A reveals a characteristic spectral shape of the pion-decay process, which indicates that the \grays\ are produced by the interactions of hadronic cosmic rays with ambient gas. The \grays\ spectrum of region B has a hard photon index of 2.12$\pm 0.02$, which is similar to other young massive star clusters. 
We argue that the diffuse GeV \gray\ emission in region A and region B likely originate from the interaction of accelerated protons in clusters with the ambient gas. 

%
%The other shows the \gray\ excess region of the Gaussian disk with a radius of $0.75^{\circ}$ at the northern part of the CNC. 
%
 %favoring the hadronic emission scenarios for the two diffuse emission components.
%spatially correlates with the dense molecular gas. Such spatial correlation favors a hadronic emission scenario although the leptonic origin cannot be ruled out.

%The regions A and B have good spatial correlation of \gray\ emission and the dense molecular gas, suggesting that two different components favors the hadronic origin. 
%We cannot rule out the leptonic origin of two regions. 
%In the hadronic scenario, 

%The origin of the \gray\ emission in this region is the interaction of the cosmic rays (CRs) accelerated by the massive binary or/and clusters with the surrounding gas clouds. 
%
%We argue that the most probable origin is the interaction of the accelerated protons in the young massive star cluster Trumpler 14, 15, 16 (Tr 14, 15, 16) and massive binary \etacarina\ with ambient gas clouds, and the total cosmic-ray (CR) proton energy is estimated to be as high as  $\sim 2.7\times10^{48}\ \rm erg$ around periastron and $\sim 3.3\times10^{48}\ \rm erg$ at astron. 
\end{abstract}

% Select between one and six entries from the list of approved keywords.
\begin{keywords}
cosmic rays – gamma-rays: ISM - open clusters and associations: individual: CNC 
\end{keywords}

%%%%%%%%%%%%%%%%%%%%%%%%%%%%%%%%%%%%%%%%%%%%%%%%%%

%%%%%%%%%%%%%%%%% BODY OF PAPER %%%%%%%%%%%%%%%%%%

%%%%%%%%%%%%%%%%% introduction %%%%%%%%%%%%%%%%%%
\section{Introduction}
The origin of cosmic rays (CRs) in the Milky Way is still a mystery. Supernova remnants (SNRs) have long been considered as the main acceleration sites of Galactic CRs \citep{1934SNRs}. Moreover, growing evidence suggests that the young massive star clusters (YMCs) play an important role in accelerating Galactic CRs. 
Several such systems have been identified, e.g., Cygnus cocoon \citep{Ackermann11, Aharonian19}, Westerlund 1 \citep{Abramowski12}, Westerlund 2 \citep{Yang18}, NGC 3603 \citep{Yang17}, 30 Dor C \citep{Abramowski15}, RSGC 1 \citep{sunRSGC1}, W40 \citep{sunw40}, Mc20 \citep{sun22}, and NGC 6618 \citep{Liu22}.

CNC is one of the most active and nearest massive star forming regions in our Galaxy. It is located in the Carina spiral arm \citep{vall14} with the distance of $\sim$2.3 kpc \citep{distance}. It contains 8 open clusters with more than 66 O-type stars, 3 Wolf-Rayet (WR) stars, and the peculiar object of the Luminous Blue Variable (LBV) \etacarina\ \citep{Smith08}. 
Potential particle acceleration sites in the CNC include massive binary systems (e.g., \etacarina ), massive star clusters (e.g., Tr 14, 15, 16), and possibly some unrecognized SNRs \citep{Smith20}. 
The central region of CNC mainly consists of the young star clusters Tr 14, 15, and 16. 
The northwestern part of CNC contains prominent ionized hydrogen (\ion{H}{ii}) region Gum 31 around the very young ($\sim$1-2 Myr) stellar cluster NGC 3324, and the oldest ($\sim$8-10 Myr) cluster NGC 3293 \citep{G22, Preibisch17}. 
%Most of the young massive star clusters in the CNC, such as Tr 14, 16, for which ages between $\sim$2 and $\sim$8 Myr have been estimated \citep{Preibisch11}
Tr 14 is one of the most extensively studied young ($\sim$1 Myr) massive clusters in our Galaxy. It contains no less than 13 O-type stars, and its total mass is estimated to be $1 \times 10^{4}$\msun\ \citep{Ascenso07}.
Tr 14 may be 1-2 Myr younger than Tr 16 \citep{Smith06}, and is closer to its associated molecular cloud than Tr 16 \citep{fujita21}. 
Tr 16 includes 42 O-type stars. It is well-known for the presence of \etacarina, which is a massive, variable star binary. \etacarina\ is composed of a primary star (a LBV star with more than $90 \msun$) and a companion star (an O or WR star with less than $30 \msun$) \citep{Hillier01, Verner05}. 
Both stars in \etacarina\ have strong winds and high mass-loss rates. The stellar mass-loss rate of the primary is $\dot{M_{\rm 1}} \approx 2.5 \times 10^{-4}\msun\ \rm yr^{-1}$ with terminal velocity of 500 km $\rm s^{-1}$ \citep{Pittard02}. The companion star has a stellar mass-loss rate of $\dot{M_{\rm 2}} \approx 10^{-5}\msun\ \rm yr^{-1}$ and faster stellar wind with  terminal velocity of 3000 km $\rm s^{-1}$ \citep{Pittard02,Parkin09}. 
In colliding-wind binaries (CWBs) or YMCs, strong shocks produced by the interaction between their powerful stellar winds which interacts with the interstellar medium  likely accelerate particles to very high energy \citep{delValle12, DeBecker13}. 
\xrays\ observations with $\it NuSTAR$ \citep{Hamaguchi18} and \grays\ observations \citep{White20} show that the accelerated particles produce non-thermal emission through colliding with surrounding gas.
%The accelerated particles produce non-thermal emission through colliding with surrounding gas based on \xrays\ observations with $\it NuSTAR$ \citep{Hamaguchi18} and \grays\ observations \citep{White20}. 

%
The Astro-Rivelatore Gamma a Immagini Leggero (\textit{AGILE}) \citep{Tavani09} and \fermi\ \citep{Abdo09} have detected \gray\ emissions from the direction of \etacarina. 
Its \gray\ spectrum measured using \fermi\ can be described by two different components with a division of 10 GeV. The high-energy component is generally suggested to be the hadronic origin \citep{Farnier11,Reitberger15,Balbo17,White20}. The origin of the lower-energy component is still uncertain %. Some people prefer leptonic origin 
\citep{Farnier11,Gupta17,Balbo17,Ohm15,White20}. 
%}, others favor hadronic origin \citep{
%
%HESS also observed this region from 2004 to 2010 to search for very high erengy (VHE) \gray\ emission from the binary and the Carina Nebula, but could only provide upper limits at energies $\geq$500 GeV \citep{HESS12}.
Recently, a point-like VHE \gray\ source from the direction of \etacarina\ was detected by HESS. Its spectrum is described by a power-law. \citep{HESS20}.

For GeV \gray\ emission, \etacarina\ is considered as a point-like source. 
%Most of them only consider this region as a point-like source from \etacarina\ to study its \gray\ emission. 
Until recently, \cite{White20} found significant extended \gray\ emission around the \etacarina, and they use a CO template of this region to model the extended emission. %and regarded it as the excesses seen in the residuals by using a CO template of the region. 
\citet{Yang18} analyzed the origin of the \gray\ emission from FGES J1036.3-5833 which includes the regions of Westerlund 2 and CNC. 
FGES J1023.3-5747 \citep{Ackermann17} and HESS J1023-575 \citep{Aharonian07,HESS11} are the diffuse \gray\ emission seen from the vicinity of Westerlund 2 \citep{Yang18,Mestre21} which seem to indicate that the YMC Westerlund 2 can provide sufficient non-thermal energy to account for the \gray\ emission.
%Hence, we consider \etacarina\ as a point source is contained in the CNC to investigated the GeV \gray\ emission. 

%
In this paper, we analyzed the \gray\ emission toward CNC taking advantage of more than 13 years of \fermi\ data, and tried to study the possible origin of CNC \gray\ emission. The paper is organized as follows. In Sect.2, we present the data set and the results of the data analysis. In Sect.3, we study the gas distributions in this region. In Sect.4, we investigate the possible origin of the \gray\ emissions. In Sect.5, the CR content around this region is discussed. Finally, In Sect.6, we discuss the implications of our results. 

%%%%%%%%%%%%%%%%% data reduction and analysis %%%%%%%%%%%%%%%%%%
\section{\fermi\ data analysis}
\label{sec:data}
We selected the latest \fermi\ Pass 8 data around the CNC region from August 4, 2008 (MET 239557417) until September 25, 2021 (MET 654297860), and used the standard LAT analysis software package $\it v11r5p3$ \footnote{\url{https://fermi.gsfc.nasa.gov/ssc/data/analysis/software/}}. 
We chose a 10\deg $\times$ 10\deg\ square region centered at the position of CNC (R.A. = 161.00$\deg$, Dec. = -59.55$\deg$) as the region of interest (ROI).
The instrument response functions (IRFs) {\it P8R3\_SOURCE\_V3} was selected to analyze the events in the ROI of evtype = 3 and evclass = 128. 
We also applied the recommended expression $\rm (DATA\_QUAL > 0) \&\& (LAT\_CONFIG == 1)$ to select the good time intervals (GTIs) based on the information provided in the spacecraft file.
In order to reduce \gray\ contamination from the Earth's albedo, only the events with zenith angles less than 90$\deg$ are included for the analysis. We used the Python module that implements a maximum likelihood optimization technique for a standard binned analysis \footnote{\url{https://fermi.gsfc.nasa.gov/ssc/data/analysis/scitools/python_tutorial.html}}.

In the background model, we included the recently released \fermi\ 10-year Source Catalog \citep[4FGL-DR2,][]{Ballet20,Abdollahi20} within the ROI enlarged by 5$\deg$. 
The source model file was generated using the script make4FGLxml.py\footnote{\url{https://fermi.gsfc.nasa.gov/ssc/data/analysis/user/}}, and all sources within 4.5$\deg$ of center were set free.
For the diffuse background components, we use the latest Galactic diffuse emission model {\it gll\_iem\_v07.fits} and isotropic extragalactic emission model {\it iso\_P8R3\_SOURCE\_V3\_v1.txt}\footnote{\url{https://fermi.gsfc.nasa.gov/ssc/data/access/lat/BackgroundModels.html}} with their normalization parameters free.

\begin{figure}
%\centering
\includegraphics[scale=0.37]{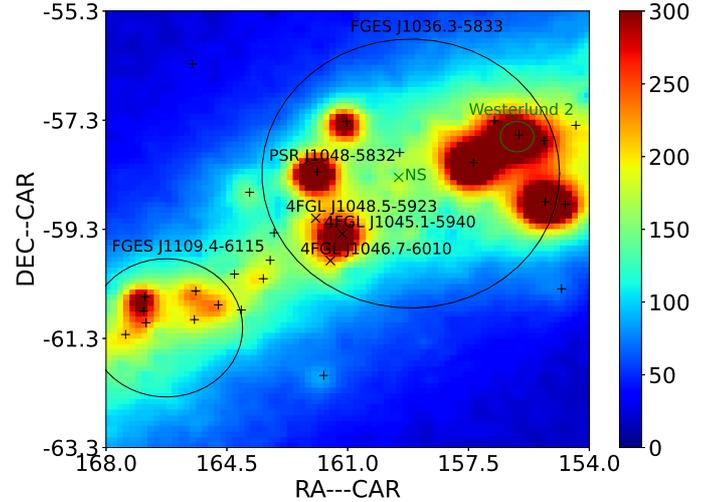}
\caption {\fermi\ counts map above 500 MeV in the $8^{\circ} \times 8^{\circ}$ region surrounding the CNC, with pixel size of $0.1^{\circ} \times 0.1^{\circ}$. All black pluses represent the 4FGL-DR2 sources within the region. The black circle shows the extended emission related to FGES J1036.4-5833. The green circle shows the YMCs Westerlund 2. The crosses indicate the three point sources 4FGL J1045.1--5940, 4FGL J1048.5-5923, 4FGL J1046.7-6010 around the CNC.}
\label{fig:cmap}
\end{figure}

First, we used the events above 500 MeV to study the spatial distribution of the \gray\ emission near CNC. The \gray\ counts map in the $8\deg \times 8\deg$ region around CNC is shown in Fig.\ref{fig:cmap}. 
In our analysis, we found a new source of which the TS and corresponding position are (TS = 129; l = 285.29; b = 0.15). It is marked with green cross in Fig.\ref{fig:cmap}.
We find no counterpart for this new source in other wavelengths.
We put the new source in the spatial analysis model with a power-law spectrum.
In addition, \citet{Mart21} found a new source (i.e. 4FGL J 1036.1-5934) identified as the nova that occurred in March 2018 \citep{Jean18}. 
This new source is also included in the latest catalog.  
Thus, to prevent contamination at the position of CNC from the emission of nova, we excluded the data from MET 542636972 to 558588527.
%We note that in the 4FGL-DR2 catalog three point sources, 4FGL J1045.1-5940 (TS=8297, $\delta=0.024^\circ$), 4FGL J1048.5-5923 (TS=146, $\delta=0.5298^\circ$), 4FGL J1046.7-6010 (TS=76, $\delta=0.8597^\circ$), are flagged as potentially associated with the massive binary \etacarina. All these three sources have the \etacarina\ within their localization error circles, it's hard to determine which source is exactly from \etacarina\ at present.
We note that there are three 4FGL-DR2 catalog point sources: 4FGL J1045.1-5940, 4FGL J1048.5-5923, 4FGL J1046.7-6010. 4FGL J1045.1-5940 was identified as the massive binary \etacarina. 
%----------------------------------------------------- FIGURE 1
\subsection{Multi-point source model}
\label{model1}
To study the excess \gray\ emission around CNC, we test several different models. The tested models are summarized in Table.\ref{table:1}.
We firstly determine a point source that used to model \etacarina\ binary stars. To do that, we excluded these three point sources from our background model. 
We added a point-like source at the $\eta$ $\rm Car's$ position into our background model, and optimized the localization using the {\it gtfindsrc} tool. 
The best-fit position of the excess above 500 MeV is [RA = $161.30\deg$ , Dec = $-59.70\deg$], with $2\sigma$ error radius of $0.15\deg$. 
In the later analysis, the point source at this position (source \etacarina) will be always included in the models and is used to represent the emission from \etacarina. 
Our first model (model 1) use three point sources (point source \etacarina, 4FGL J1046.7-6010 and 4FGL J1048.5-5923) to model the excess \gray\ emission around CNC. 
Each point source has a LogParabola spectral shape. 
We performed a binned likelihood analysis to derive the likelihood value ($-\log({\cal L})$) and the Akaike information criterion (AIC, \cite{Akaike1974}) value. The AIC is defined as AIC = $-2\log({\cal L}) + 2k$, where $k$ is the number of free parameters in the model. 
%We performed a binned likelihood analysis. 
%The derived likelihood values ($-\log({\cal L})$) and the Akaike information criterion (AIC, \cite{Akaike1974}) values for the multi-point sources model are -1265307 and -2530414, respectively.
%The AIC is defined as AIC = $-2\log({\cal L}) + 2k$, where $k$ is the number of free parameters in the model.
The derived $-\log({\cal L})$ and AIC for the multi-point source model are -1265307 and -2530414, respectively.

\subsection{Spatial template for two Gaussian disks}
\label{sec:two Gaussian disks}
To further investigate the diffusion of the GeV \gray\ emission, we used a spatial template that consists of two regions (A, B). Each region are modelled as a Gaussian disk and we varied the positions and sizes of the disks to find the best-fit parameters. 
The significance of the extended source is quantified by $\rm TS_{ext}=2\log({\cal L}_{ext} /{\cal L}_{ps})$, where $\rm {\cal L}_{ext}$ is the maximum likelihood for the extended source model, and $\rm {\cal L}_{ps}$ for the point-like sources model \citep{Lande12}. 
For region A, we found that the \gray\ emission of \etacarina\ is very strong. 
Hence, the added the center of Gaussian disk is set at above best-fit position. The radius of the disk varies from $0.2\deg$ to $1\deg$ in steps of $0.05\deg$. 
%Hence, we added the Gaussian disk centered at the above best-fit position, vary their radius from $0.2\deg$ to $1\deg$ in steps of $0.05\deg$. 
We used this Gaussian disks to replace the spatial components of the two unassociated point sources: 4FGL J1046.7-6010 and 4FGL J1048.5-5923 in the multi-point source model. 
The likelihood ratio peak at the Gaussian disk template with a radius of $0.4\deg \pm\ 0.02\deg$ can fit the \gray\ excess for the central part of the CNC (region A). 

\begin{figure}
%\centering
\includegraphics[scale=0.37]{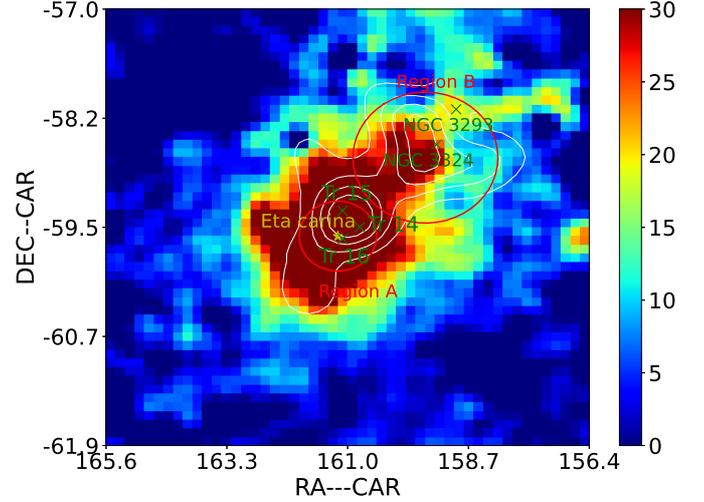}
\caption {Residual map above 500 MeV near the CNC after subtracting two Gaussian disks with a radii of 0.4$\deg$ and 0.75$\deg$ from the model 2. Details are given in Sect.~\ref{sec:two Gaussian disks}. 
The map has a size of $5\deg \times\ 5\deg$ ($0.1\deg \times\ 0.1\deg$ pixelsize) and has been smoothed with a Gaussian kernel of 0.9$\deg$. 
The regions A and B is marked with the red circles. The green crosses indicate the star clusters. The yellow asterisk indicates the position of \etacarina. The white contours show the $\rm H_2$ column density derived in Sect.~\ref{sec:Gas}. }
\label{fig:residual}
\end{figure}

It was found that there exists extended residual \gray\ emission in the northwestern part of the CNC (region B) after this process. 
Therefore, to find out whether the residual emission is extended or not, we used a point-like source or a Gaussian disk to model this residuals. 
The position of the added point source or the center of the Gaussian disk is set to the peak position of the residuals.
The tested radius of the Gaussian disk varies from $0.2\deg$ to $1\deg$ with a step of 0.05 for region B. 
In the above test, we find that the Gaussian disks with the radii of 0.75$\deg$ for the region B and 0.4$\deg$ for the region A (model 2) can best fit the data. 
The derived -$\log({\cal L})$ and AIC for this model are -1265580 and -2530978 , respectively. 
We subtracted the $0.4\deg$ Gaussian disk and the $0.75\deg$ Gaussian disk from the model 2 and derived the residual map shown in Fig.\ref{fig:residual}. 
The map revealed significant diffuse residuals in this region. 
In the following subsections, we test different spatial models of these diffuse emissions. 

%----------------------------------------------------- FIGURE 2
\subsection{Spatial template for molecular hydrogen}
\label{sec:CO model}
To determine whether the extended GeV \gray\ emission is correlated with the gas distribution, we considered a spatial template of H$_2$. 
For the H$_2$ template, we use the Carbon monoxide (CO) composite survey (see Sect.~\ref{sec:Gas} for details) to produce. 
As shown in Fig.\ref{fig:residual}, the white contours represent the H$_2$ distribution from the CO measurements, which overlap well with the excess emission. 
We added the H$_2$ template (with a power-law spectrum) into the multi-point source model to obtain our model 3. 
Through the binned likelihood analysis, the derived -$\log({\cal L})$ and the AIC for the H$_2$ template are  -1265540 and -2530890, respectively. 

\subsection{Spatial template for molecular and ionized hydrogen}
\label{sec:CO+HII model}
CNC contains one of the largest and most active \ion{H}{II} regions in our Galaxy. The derivation for the details of the \ion{H}{II} distribution is given in Sect.~\ref{sec:Gas}. 
We also note that the \gray\ emission has good spatially correlation with the \ion{H}{II} gas (see Sect.~\ref{sec:Gas}).
Thus, we adopted a spatial template considering \ion{H}{II} + H$_2$ gases. 
%Details of the \ion{H}{II} template see Sect.~\ref{sec:Gas}. 
%The \ion{H}{II} template is produced using the $Planck$ free-free map (see Sect.~\ref{sec:Gas}). 
We summed the column density of \ion{H}{II} and H$_2$ gases to generate the H$_2$ + \ion{H}{II} template with a power-law spectral shape. 
The H$_2$ + \ion{H}{II} template is added into the multi-point source model as a diffuse component (model 4).
%The H$_2$ + \ion{H}{II} template as a diffuse component to added the multi-point sources model (model 4). 
After performing the binned likelihood analysis, the derived -$\log({\cal L})$ and the AIC for the spatial model of H$_2$ + \ion{H}{II} are -1265619 and -2531048, respectively.

%----------------------------------------------------- TABLE 1
\begin{table*}
%\centering
        \caption{Spatial analysis (>500MeV) results for the different models.}
\begin{tabular}{lcccc}
\hline
\hline
        Model & -$\log({\cal L})$ & $\rm TS_{ext}$ & d.o.f. & $\rm \Delta AIC$ \\

\hline
        Model 1 (multi-point sources)& -1265277 & - & 93 & - \\ %\cline{1-1}
        Model 2 ($0.4\deg$ Gaussian disk + $0.75\deg$ Gaussian disk)& -1265580 & 606 & 91 & -610 \\ %\cline{1-1}
        Model 3 (multi-point sources + $\rm H_2$ template) & -1265540 & 526 & 95 & -522 \\ %\cline{1-1}
        Model 4 (multi-point sources + $\rm H_2$ + \ion{H}{ii} template) & -1265619 & 684 & 95 & -680 \\ %\cline{1-1}
        Model 5 (multi-point sources + $0.4\deg$ Gaussian disk + $0.75\deg$ Gaussian disk)& -1265647 & 740 & 97 & -732 \\ %\cline{1-1}
\hline
\hline
\end{tabular}
\label{table:1}
\end{table*}

\subsection{Spatial template for two Gaussian disks and multi-point sources}
In the model 2, we removed the two point sources 4FGL J1046.7-6010 and 4FGL J1048.5-5923 from the background and accounted for the emissions close to CNC with extended components. Here we further test the inclusion of both the extended components and two point sources in the model (model 5). %We note that the background point sources 4FGL J1046.7-6010 and 4FGL J1048.5-5923 are located close to CNC.  As shown in Fig.\ref{fig:residual}, we find another peaks in the residual map. 
Thus, we added the two 4FGL point sources back into the model file based on the model 2. Each point source has a LogParabola spectral shape. 
We find that this model (multi-point sources + $0.4\deg$ + $0.75\deg$ Gaussian disks), can explain the observational data best among the 5 models we have considered. 
The derived -$\log({\cal L})$ for model 5 is equal to -1265647, and the AIC is equal to -2531100. 

To compare the goodness of the fit in the different models, we also calculated the $\rm \Delta AIC$, the AIC of the model 1 and the model 2-5.
It is evident from Table \ref{table:1} that the model with the two Gaussian disks and multi-point sources provides the highest $\rm TS_{ext}$ value and the minimum $\rm \Delta AIC$ value. 
Therefore, in the following analysis, we use the model 5 as the spatial template. 

The derived photon index for $0.4\deg$ Gaussian disk above 500 MeV is $2.36 \pm 0.01$ and the total \gray\ flux can be estimated as $(2.65 \pm 0.02) \times 10^{-8} \rm ph\ cm^{-2}\ s^{-1}$. 
Considering the distance of about 2.3 kpc, the total \gray\ luminosity is estimated to be $(1.34 \pm 0.01) \times 10^{34} \rm erg\ s^{-1}$ above 500 MeV with the single power-law spectrum. 
The photon index of 0.75$\deg$ Gaussian disk is $2.12 \pm 0.02$ and the total flux is estimated as $(1.97 \pm 0.02) \times 10^{-8} \rm ph\ cm^{-2}\ s^{-1}$, corresponding to $(9.95 \pm 0.08) \times 10^{33} \rm erg\ s^{-1}$ above 500 MeV. 
We found that this model give the best log-likelihood value. We argue these additional point sources may represent the contamination from bright central source \etacarina. In our following spectral analysis we derive the spectral energy distributions (SEDs) of the two Gaussian disks with and without the additional multiple point sources (model 2 and model 5), respectively. We found the results are consistent with each other within error bars. And we include the difference as systematic errors.

\subsection{Spectral analyses}
\label{sec:spectral_analy}
\begin{figure}
%\centering
\includegraphics[scale=0.41]{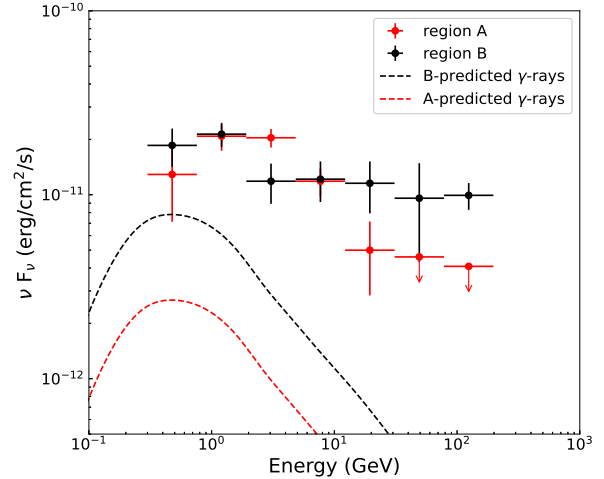}
\caption {SEDs of \gray\ emission in the region A (black point) and region B (red data) based on the model 5. The dashed curves represent the predicted \gray\ emission assuming the CR density in the two regions is the same as those measured locally by AMS-02 \citep{Aguilar15}. Both the statistical and systematic errors are considered. For details, see the context in Sect.~\ref{sec:spectral_analy}.}
\label{fig:sed}
\end{figure}

We used the best-fit spatial template as the spatial model of the extended \gray\ emission, and assumed a power-law spectral shape to extract the SED. 
We divided the energy range 300 MeV-200 GeV into seven logarithmically spaced energy bins, and in each bin the SED flux is derived via the maximum-likelihood method. 
%The energy bins are all detected with a significance of less than 2$\sigma$. 
We calculated the upper limits within 3$\sigma$ for the energy bins with a significance lower than 2$\sigma$. 
Fig.\ref{fig:sed} shows the derived SEDs of the regions A (black) and B (red). 
The dashed line represents the predicted \gray\ emissions assuming the CR density in regions A and B, respectively. 
%We further divided the \gray\ flux by the average gas column density in Sect.~\ref{sec:Gas} to get the \gray\ emissivity per H atom. 
%Assuming the \gray\ are produced by interaction between CRs and ambient gas, the \gray\ emissivity per H atom should be proportional to the CR density. 
In the analysis, we estimated the uncertainties of SEDs system due to the Galactic diffuse emission model and the LAT effective area ($\rm A_{eff}$) by changing the normalization by $\pm 6\%$ from the best-fit value for each energy bins, and considered the maximum flux deviations of the source as the systematic error \citep{Abdo09a}. 

%----------------------------------------------------- FIGURE 3

\section{Gas content around CNC}
\label{sec:Gas}
We investigated three different gas phases, i.e., the H$_{2}$, the neutral atomic hydrogen (\ion{H}{i}), and the \ion{H}{ii}, in the vicinity of CNC region. 

\begin{figure*}
%\centering
\includegraphics[scale=0.24]{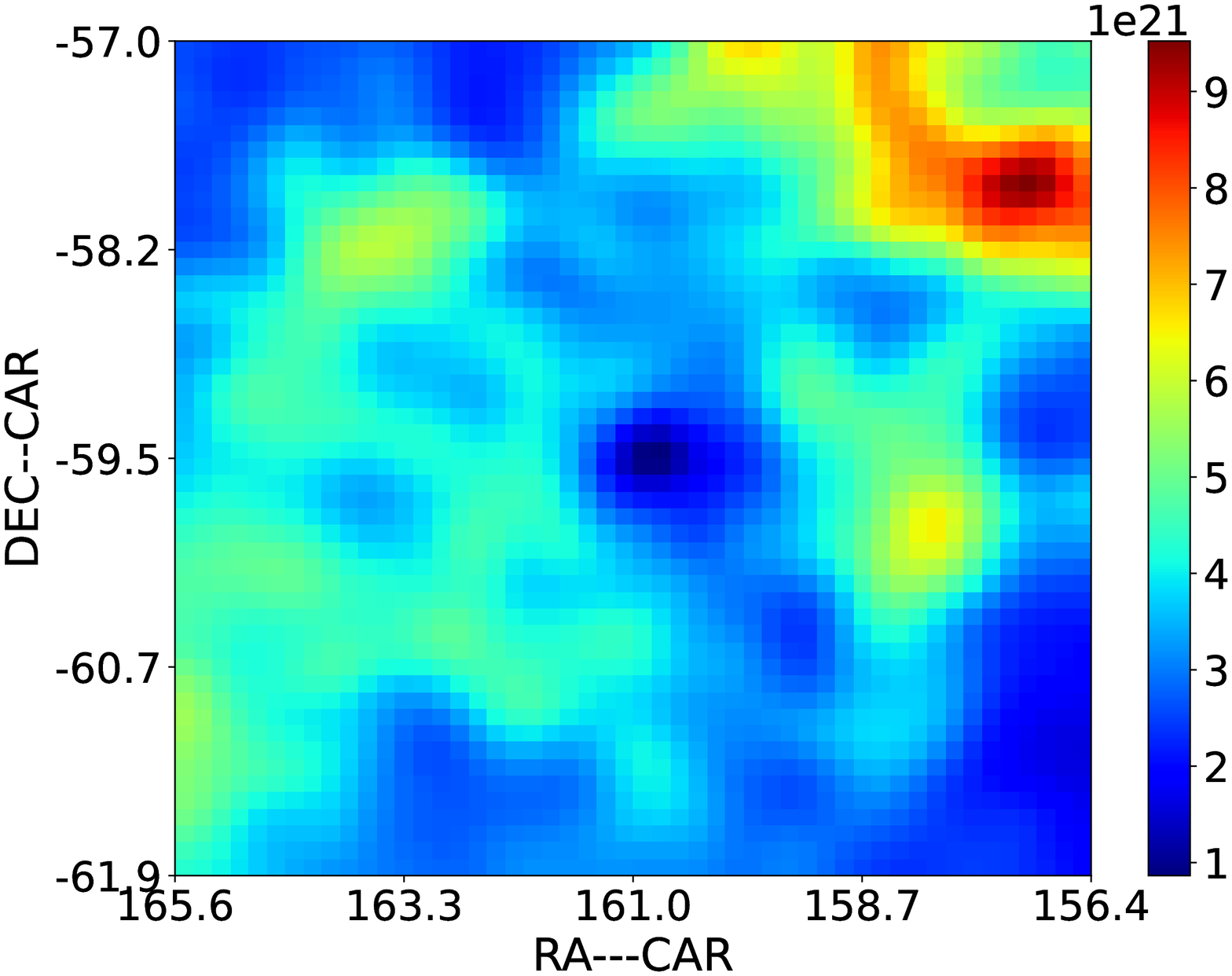}
\includegraphics[scale=0.24]{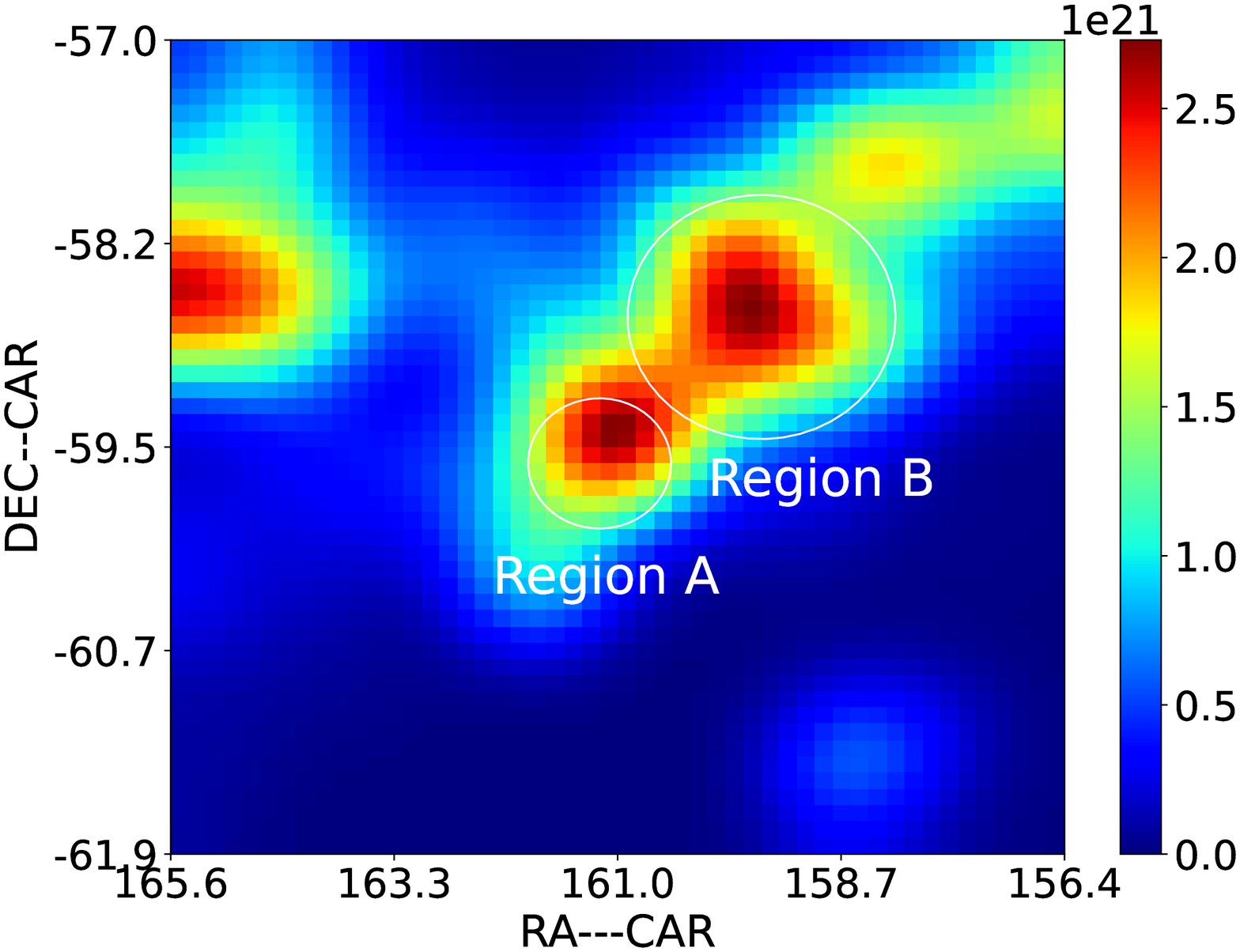}
\includegraphics[scale=0.24]{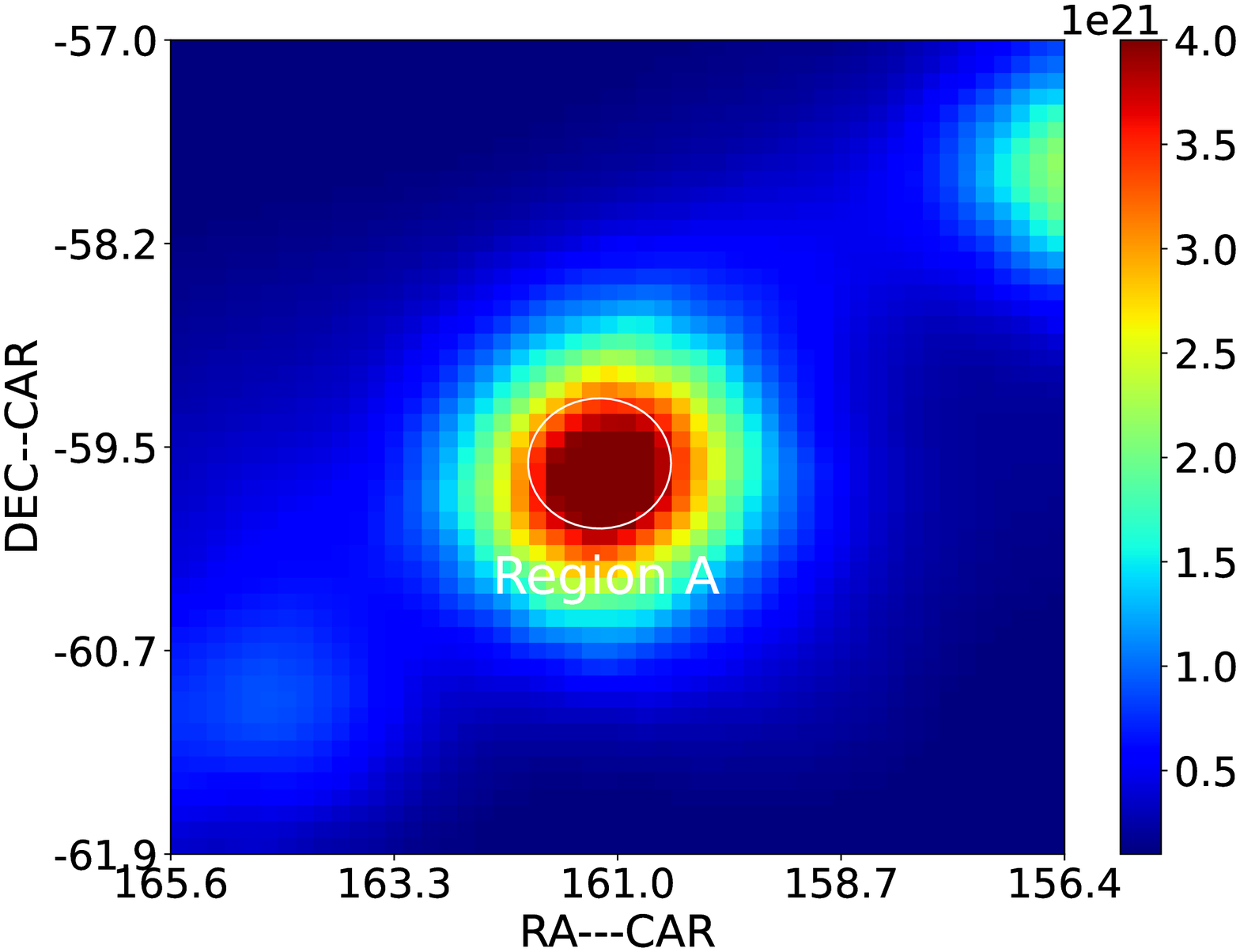}
\caption {Maps of gas column densities in three gas phases. Left shows the map of \ion{H}{i} column density derived from 21-cm all-sky survey. Middle shows the H$_{2}$ column density derived from the CO data. Right shows the \ion{H}{ii} column density derived from the \planck\ free-free map assuming the effective density of electrons $n_{\rm e}=10~\rm cm^{-3}$. The white circles indicate the region A and region B, which is the same as the red circles in Fig.\ref{fig:residual}. For details, see the context in Sect.\ref{sec:Gas}.}
\label{fig:Gas}
\end{figure*}

The \ion{H}{i} data is from the data-cube of the \ion{H}{i} $\rm{4\pi}$ survey (HI4PI), which is a 21-cm all-sky database of Galactic \ion{H}{i} \citep{HI4PI16}. 
We estimated the \ion{H}{i} column density using the equation,
\begin{equation}
N_{\ion{H}{i}} = -1.83 \times 10^{18}T_{\rm s}\int \mathrm{d}v\ {\rm ln} \left(1-\frac{T_{\rm B}}{T_{\rm s}-T_{\rm bg}}\right),
\end{equation}
where $T_{\rm bg} \approx 2.66\ \rm K$ is the brightness temperature of the cosmic microwave background radiation at 21 cm, and $T_{\rm B}$ is the brightness temperature of the \ion{H}{i} emission. 
In the case when $T_{\rm B} > T_{\rm s} - 5\ \rm K$, we truncate $T_{\rm B}$ to $T_{s} - 5\ \rm K$; $T_{s}$ is chosen to be 150 K. 
The derived \ion{H}{I} column map integrated in the velocity range $v_{\rm LSR}=[-32,-5]$ $\rm km\ s^{-1}$ \citep{Seo19,Rebolledo21} is shown in the left panel of Fig.\ref{fig:Gas}. 
We also use this range to integrate the line emission of the CO in this velocity range. 

We use the CO composite survey \citep{Dame01} to trace the $\rm H_{2}$. 
The standard assumption of a linear relationship between the velocity-integrated brightness temperature of CO 2.6-mm line, $W_{\rm CO}$, and the column density of molecular hydrogen, $N(\rm H_{2})$, i.e., $N({\rm H_{2}}) = X_{\rm CO} \times W_{\rm CO}$ \citep{Lebrun1983}. 
$X_{\rm CO}$ is the $\rm H_{2} / CO$ conversion factor that chosen to be $\rm 2.0 \times 10^{20}\ cm^{-2}\ K^{-1}\ km^{-1}\ s$ as suggested by \cite{Dame01} and \cite{Bolatto13}. 
The derived molecular gas column density is shown in the middle panel of Fig.\ref{fig:Gas}.

%----------------------------------------------------- FIGURE 4
CNC is one of the largest and most active \ion{H}{ii} regions in the Galaxy. 
To obtain the \ion{H}{ii} column density we used the \planck\ free-free map \citep{Planck16}. 
First, we transformed the emission measure (EM) into free-free intensity by using the conversion factor in Table 1 of \cite{Finkbeiner03}. 
Then, we calculate the \ion{H}{ii} column density from the intensity ($I_{\nu}$) of free-free emission by using Eq.(5) of \cite{Sodroski97}, 
\begin{equation}
\begin{aligned}
  N_{\ion{H}{ii}} = &1.2 \times 10^{15}\ {\rm cm^{-2}} \left(\frac{T_{\rm e}}{1\ \rm K}\right)^{0.35} \left(\frac{\nu}{1\ \rm GHz}\right)^{0.1}\left(\frac{n_{\rm e}}{1\ \rm cm^{-3}}\right)^{-1} \\
&\times \frac{I_{\nu}}{1\ \rm Jy\ sr^{-1}},
\end{aligned}
\end{equation}
where $\nu = \rm 353\ GHz$ is the frequency, and an electron temperature of $T_{e} =\rm 8000\ K$. 
The \ion{H}{ii} column density is inversely proportional to the effective density of electrons $n_{\rm e}$. 
Thus, we adopted an effective density $10\ \rm cm^{-3}$, which is the value suggested in \cite{Sodroski97} for the region inside the solar circle. 
The derived \ion{H}{ii} column density is also shown in the right panel of Fig.\ref{fig:Gas}. 
We note that the \ion{H}{ii} gas distribution is similar to that of the YMCs NGC 3603 \citep{Yang17} and W40 \citep{sunw40}. 
%They all have that the \gray\ emission region shows good spatial consistency with the \ion{H}{ii} column density, which supports the hypothesis that the \gray\ emission comes from the interaction of the accelerated particles with surrounding gas in the supper bubble. 
They all have the \gray\ emission region, which shows good spatial consistency with the \ion{H}{ii} column density. 
Moreover, several YMCs are located here, i.e Tr 16, Tr 14, which had ionized the ambient media in the CNC. 

%----------------------------------------------------- TABLE 2
\begin{table}
%\centering
        \caption{Gas total masses and number densities within the region A and region B. See Sect.~\ref{sec:Gas} for details.}
\begin{tabular}{lcccc}
\hline
        Tracer & Region & Mass ($\rm{10^{4}\msun}$) & Number density ($\rm {cm^{-3}}$) \\
\hline
        H$_{2}$ + \ion{H}{II} & A & 5.54 + 4.31 & 231 \\ %\cline{1-1}
        H$_{2}$ + \ion{H}{II} & B & 16.18 + 3.99 & 72 \\ %\cline{1-1}
\hline
\end{tabular}\\
\label{table:2}
\end{table}
The total mass within the cloud in each pixel can be calculated from the expression
\begin{equation}
M_{\rm H} = m_{\rm H} N_{\rm H} A_{\rm angular} d^{2}
\end{equation}
where $M_{\rm H}$ is the mass of the hydrogen atom, $N_{\rm H} = N_{\ion{H}{ii}} + 2N_{\rm H_{2}} + N_{\ion{H}{i}}$ is the total column density of the hydrogen atom in each pixel. $A_{\rm angular}$ is the angular area, and $d$ is the distance of CNC. 
The total mass in the GeV \gray\ emission region is estimated to be $\sim 5.54 \times 10^{4}~\msun$ for region A and $\sim 1.62 \times 10^{5}~\msun$ for region B.
If we assume the GeV \gray\ emission within both the regions A and B are spherical in geometry, with the corresponding sizes of $0.4\deg$ and $0.75\deg$. 
The radius can be estimated as $r_{\rm{A,B}} = d \times \theta_{\rm{A,B}} (\rm{rad})$, and $d$ is the distance to the objective region. 
The averaged over the volume gas number density of region A is $\rm n_{gas}=231\ cm^{-3}$, while for region B, the values is $\rm n_{gas}=72\ cm^{-3}$. 
Table \ref{table:2} shows the gas total masses and number densities within the region A and region B.

\section{The origin of gamma-ray emission}
%\label{sec:origin}
%
There is the massive binary \etacarina\ in the center of region A.
In spatial analysis, we added \etacarina\ in the background model as a point-like source. 
PSR J1048-5832 is located about 1.2$\deg$ away from the center of the \gray\ emission region (as shown in Fig.\ref{fig:cmap}), which makes the association of the diffuse \gray\ emission to this pulsar unlikely \citep{Danilenko13}. 
%The region of PSR J1048-5832 has a distance of 2.7 kpc and a spin-down luminosity of $2 \times 10^{36}\ \rm erg\ \rm s^{-1}$. There is no pulsar in this large distance that have enough spin-down power to provide the $1.7 \times 10^{36}\ \rm erg\ \rm s^{-1}$ \gray\ luminosity. 
%This makes the association of the diffuse \gray\ emission to the pulsar unlikely. 
On the other hand, there are no known SNRs inside this region. 

\subsection{Region A}
\label{sed1}
\begin{figure}
%\centering
\includegraphics[scale=0.42]{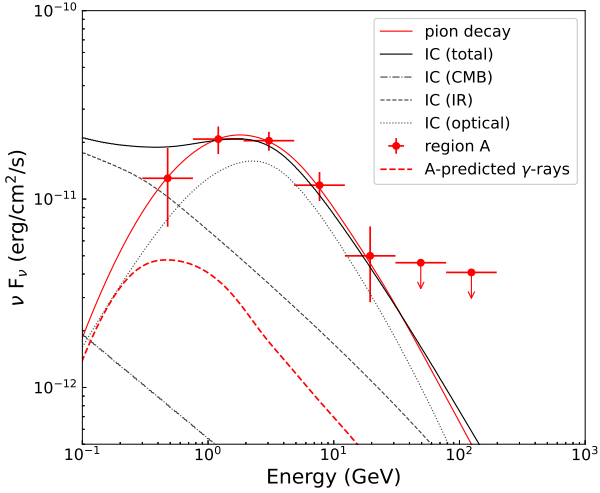}
\caption {SED of emission in the region A for a $0.4\deg$ Gaussian disk spatial model. The red dashed line represents the predicted \gray\ emissions assuming the CR density in this region is the same as those measured locally by AMS-0.2 \citep{Aguilar15}.}
\label{fig:sed1}
\end{figure}

\cite{White20} found that the significant extended \gray\ emission in this region, suggesting an additional component associated with CRs interactions. 
%Furthermore, two young massive star clusters (Tr 14, 16) in the CNC are another natural acceleration site of the CRs. 
%Thus, the CRs accelerated by those clusters interacting with ambient gas provides a explanation of the diffuse \gray\ emissions. 
In our case, the YMCs Tr 14 and Tr 16 are the most promising sources of the CRs. 
Although we cannot rule out the possibility that \etacarina\ is the source of the CRs that produced the diffuse \gray\ emissions, we postulate here that those clusters may be another natural acceleration site of the CRs. 
We used Naima\footnote{\url{https://naima.readthedocs.io/en/latest/index.html}} \citep{Zabalza15} to fit the SEDs. 
Naima is a numerical package that allows us to implement different functions and includes tools to perform Markov chain Monte Carlo (MCMC) fitting of nonthermal radiative processes to the data. 
We note that the extended GeV emission and the molecular hydrogen gas are spatially correlated. 
Thus, we assume the \grays\ are produced in the pion-decay process from the interaction of the CRs with the ambient molecular clouds. 
Because the low-energy data points are poorly constrained, we used a broken power-law spectrum 
\begin{equation}
N(E) =\begin{cases} A(E/E_0)^{-\alpha_1} & \mbox{: }E<E_\mathrm{b} \\ A(E_\mathrm{b}/E_0)^{(\alpha_2-\alpha_1)}(E/E_0)^{-\alpha_2} & \mbox{: }E>E_\mathrm{b} \end{cases},
\label{equ:bpl}
\end{equation}
to fit the GeV \gray\ data. 
The derived SED is shown in Fig.\ref{fig:sed1}.
%This is shown by the red points in Fig.\ref{fig:sed1}. 
We treated A, $\alpha_1$, $\alpha_2$, $E_\mathrm{b}$ as free parameters for the fitting. 
As calculated in this section, the average number densities of the target protons for regions A and B are 231 $\rm cm^{-3}$ and 72 $\rm cm^{-3}$, respectively, which are derived from the gas distributions in Sect.~\ref{sec:Gas} consider $\rm H_2$ + \ion{H}{II}. 
In Fig.\ref{fig:sed1}, we present the best-fitting results for region A. The maximum log-likelihood value is -0.1. The derived parameters are $\alpha_1 = 0.9 \pm 0.3$, $\alpha_2 = 3.2^{+0.7}_{-0.3}$, $E_\mathrm{b} = 21^{+4}_{-3} \rm GeV$ and the total energy is $W_{\rm p} = (1.11 \pm 0.12) \times 10^{48}\ \rm erg$ for the protons above 2 GeV. 
The red dashed line in Fig.\ref{fig:sed1} represents the predicted the fluxes of \gray\ emissions based on the $\rm H_2$ + \ion{H}{II} column density map in region A, assuming the CRs have the same as the local measurement by AMS-02 \citep{Aguilar15}. 
We found that a significant CR enhancement in this region, which is predicted near the CR acceleration site. 
%----------------------------------------------------- FIGURE 5

We also tested the leptonic scenario that the \grays\ are generated via the inverse Compton (IC) scattering of relativistic electrons off the low-energy seed photons around this region. 
%We have also considered the interstellar radiation field (ISRFs) of the IC calculations}
For the photon field of the IC calculations, we considered the cosmic microwave background (CMB) radiation field, optical-UV radiation field from the star light, and the dust infrared radiation field based on the model by \cite{Popescu17}.
%For the interstellar radiation field of the IC calculations, we include the optical and UV fields from the star light, the infrared fields from the dust emission, and the cosmic microwave background (CMB) . 
We note that region A is located in \ion{H}{II} regions, the ionizing massive stars will increase the optical and UV fields significantly and thus produce additional IC emissions \citep{LiuYang22}.
%In the \ion{H}{II} regions, the ionizing massive stars will increase the optical and UV fields significantly and thus produce additional IC emissions.
%the cosmic microwave background (CMB), infrared, and optical emission calculated by \cite{Popescu17}.
We calculated the IC spectrum using the formalism described in \cite{Khangulyan14}. 
To fit the lower energy break in the \gray\ spectrum, we require a relevant break in the spectrum of parent electrons. 
Thus, we assumed a broken power-law distribution of the relativistic electrons. 
As is shown in Fig.\ref{fig:sed1}, the solid black curve represents the total predicted \gray\ emission from the IC upscattering of the seed photons by relativistic electrons. 
The derived parameters for the electrons are $\alpha_1 = 0.51 \pm 0.1$, $\alpha_2 = 4.1^{+0.6}_{-0.4}$, $E_\mathrm{b} = 15.3^{+3.0}_{-1.7}\ \rm GeV$ and the total energy of the electrons (>2 GeV) is $W_e = (7.8 \pm1.2) \times 10^{49}\ \rm erg$. 
The IC model can fit the observable data well, and the corresponding maximum-likelihood values is -0.91. 
We cannot formally rule out the leptonic origin of this region.

\subsection{Region B}
\begin{figure}
%\centering
\includegraphics[scale=0.42]{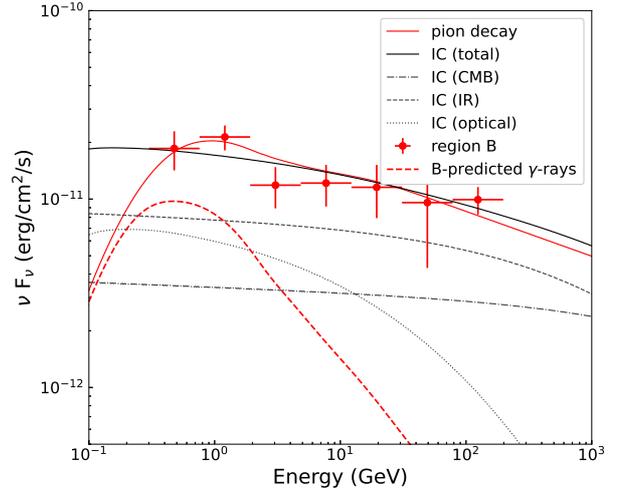}
\caption {SED of emission in the region B for a $0.75\deg$ Gaussian disk spatial model. The red dashed line represents the predicted \gray\ emissions assuming the CR density in this region is the same as those measured locally by AMS-0.2 \citep{Aguilar15}.}
\label{fig:sed2}
\end{figure}

It should be noted that there is another possible particle accelerator located at the northwestern part of the CNC. 
For region B, including a bubble-shaped young \ion{H}{ii} region Gum 31 around the young stellar clusters NGC 3324, and NGC 3293 \citep{G22}. 
Due to the good spatial correlation of the extended GeV \gray\ emission and the $\rm H_2$ gas, it is very probable that the \grays\ are related to the $\rm H_2$ gas. 
Thus, we used a hadronic scenario in which high energy \grays\ produced in the pion-decay process follow the proton-proton inelastic interactions, using the parameterization of the cross-section of \cite{Kafexhiu14}.
In this region, we used a single power-law spectrum for the parent proton distribution,
\begin{equation}
        N(E) = A~E^{-\alpha},
\label{equ:pl}
\end{equation} 
treating $A$, $\alpha$ as free parameters for the fitting. 
As shown in Fig.\ref{fig:sed2}, we present the best-fit results for region B. The maximum log-likelihood value is -1.87.
The derived index of the region B, $\alpha = 2.27 \pm 0.05$, with the total energy $W_{\rm p} = (5.0 \pm 0.4) \times 10^{48}\ \rm erg$ for the protons above 2 GeV. 

In addition to the hadronic model, we also tried to consider a leptonic model where the \grays\ are from the IC scattering. 
We assume that the electrons have a single power-law spectrum. The target photon fields for relativistic electrons to scatter include the CMB, infrared, and optical field adopted from the local interstellar radiation field calculated in \cite{Popescu17}. As shown in Fig.\ref{fig:sed2}, the leptonic model can also explain the observed \gray\ emission with a maximum log-likelihood value of -2.17.
%We cannot rule out the IC origin of the \gray\ emissions from the region B. 
The derived index of the electrons is $\alpha = 3.05 \pm 0.07$. However, for the IC model the derived energy budget for the relativistic electrons (>2 GeV) is as high as $(1.1 \pm 0.3) \times 10^{50}\ \rm erg$, which is almost $10\%$ of the typical kinetic energy of a supernova explosion $10^{51} \rm erg$. %If We change the minimum electrons energy to 10 GeV, the total energy of the electrons (>10 GeV) is $(2.4 \pm 0.5) \times 10^{49}\ \rm erg$. 
The total kinetic energy supplied by the stellar wind from a single massive star within $\sim$1 Myr, is $\sim 3 \times 10^{48}\ \rm erg$ \citep{Ezoe06}. Since CNC contains > 66 OB stars, the stellar wind power produced by the CNC $>2\times 10^{50} \rm erg$. If we consider the different clusters i.e. Tr 14 (20 OB stars), Tr 16 (21 OB stars) \citep{Shull21} and NGC 3324 (20 OB stars) \citep{Bisht21}, we evaluate the stellar wind power $6 \times 10^{49}\ \rm erg$, $6.3 \times 10^{49}\ \rm erg$ and $6 \times 10^{49}\ \rm erg$, respectively.

%----------------------------------------------------- FIGURE 6

\section{CR content in the vicinity of the CNC}
\label{sec:CR}

%Recently, revealed $1/r$ profiles in the vicinity of several YMCs indicate a continuous injection of CR protons from the central source to surroundings \citep{Aharonian19,Yang22}.
%To study the propagated CR contents in the vicinity, rather than divide the \gray\ emission region into rings as in \citet{sunw40}, we divided it into two Gaussian disks. 
%This is because the potential accelerator, the young star cluster NGC 3324, is far from the center of the \gray\ emission region. 
%
The spatial distributions of CRs can provide key information about their injection history.
In Westerlund 1, Cygnus Cocoon \citep{Aharonian19}, and Westerlund 2 \citep{Yang18}, the 1/r CR profiles are derived, which implies the continuous injection and diffusion dominated propagation of CRs from these massive star clusters \citep{Yang22}.

\begin{figure}
%/centering
\includegraphics[scale=0.42]{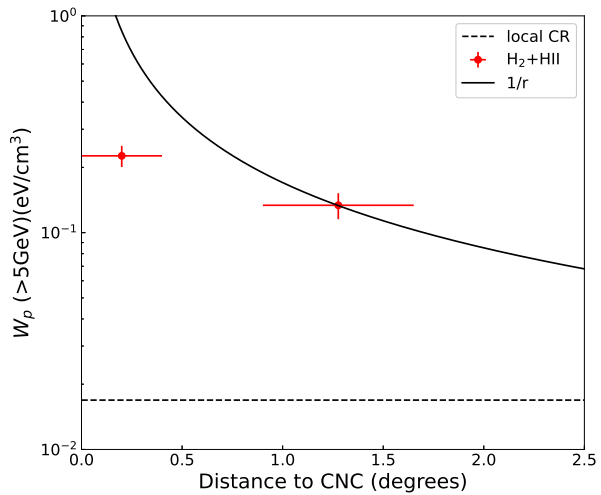}
\caption {Derived CR density profile near the CNC. The data points are the \gray\ emission above 5 GeV of the CNC. The solid curve is the 1/r profile, which is predicted by the continuous injection. For details, see the context in Sect.~\ref{sec:CR}.}
\label{fig:CR}
\end{figure}

We study the propagated CR contents in the vicinity of CNC. %we derived the CR density distributions of the regions A and B as shown in Fig~\ref{fig:CR}, respectively. 
Because of the limited angular resolution and size of this region, we chose only region A and region B, rather than annuli to derive the CR density. In both regions, the gamma-ray flux above 5 GeV and gas mass are derived separately.
%We performed a likelihood analysis for the two regions, and derived the \gray\ luminosity above 5 GeV. %This is because the inhomogeneity of the diffuse \gray\ emission and the existence of the potential accelerator in both regions. 
For gas content, the results of spatial analysis in Sect.~\ref{sec:CO+HII model} have shown that the \gray\ emission region has good spatial consistency with the H$_2$ + \ion{H}{ii} template. Thus, we derived the corresponding gas mass for regions A and B, respectively.
Then we calculated CR density according to the function (A.4) in \citet{Huber13}.
%
%We then calculated the content of CR density by using the total \gray\ luminosity \citep{Huber13}.
The derived CR density from the two \gray\ emission regions and gas distributions are shown in Fig.~\ref{fig:CR}. The first red point is related to region A and the second one is related to region B.
%\begin{equation}
%L_{\gamma} = 2 n_{\rm gas} n_{\rm CR} v_{\rm CR} \sigma_{\rm pp} E_{\gamma}\ .
%\end{equation}
%Here,  $n_{\rm CR}$ and $v_{\rm CR}$ are the number density and velocity of the CRs, respectively, and $n_{\rm gas}$ is the ambient gas density. $E_{\gamma} = \frac{2}{\kappa}E_{\rm CR}$ is the photons energy, where $\kappa \sim 0.17$ is the fraction of energy transferred from the proton to pion \citep{Kelner06}.
%
%Using the \gray\ production cross-section \citep{Kafexhiu14}, we derived the CR density from the two \gray\ emission regions and gas distributions, the results are shown in Fig.~\ref{fig:CR}. 
As mentioned in \citet{Aharonian19}, the radial distribution of the CR density in the form,
\begin{equation}
 w(r)=  w_0 (r/r_0)^{-1} \ 
\label{w(r)}
\end{equation}
where r$_0$ is assumed to be 10 pc, i.e. normalise the CR proton density $w_0$ outside but not  far from the cluster.
The total energy of CR protons within the volume of the radius $R_0$ is 
\begin{equation} 
W_{\rm p}= 4 \pi \int_0^{R_{\rm 0}} w(r) r^2  \,\mathrm{d}r  \approx \\ 2.7 \times 10^{47} (w_0/1 \ \rm eV/cm^3) (R_0/10 \ \rm pc)^2 \ \rm erg \ .
\label{Wcr}
\end{equation}
The derived CR profile is shown in the Fig.~\ref{fig:CR}. We fit the density profile using a 1/r type distribution, which would result from a continuous injection of CRs from region A.
We found no 1/r dependence and the CR density of region B is significantly lower than that of region A. 
For such a profile, the CR content in this system are indeed from some recent impulsive process rather than continuous injections. 
Certainly, here we note the recent studies have shown that the CR radial profiles for Westerlund 1 \citep{Aharonian22} and Cygnus Cocoon \citep{Abeysekara21} are not compatible with 1/r. Moreover, from a theoretical point of view, the 1/r profile is not the only possible outcome for the case of a steady central source. Actually, flatter spatial profiles are more favoured in the case of acceleration by the termination shock of stellar winds \citep{Morlino21}.

\section{Discussion and conclusion}
\label{sec:conc}

%----------------------------------------------------- FIGURE 7

%conclusion
In this paper we analyze the \grays\ emission from the massive star forming region of CNC. 
We found that the \gray\ emission in this region can be resolved into three components. 
Besides the point-like source coincides with the massive binary \etacarina,
we further found two diffuse \gray\ emission components that are spatially correlated with the molecular and ionized gas. 

The \gray\ emission from the CNC (region A) can be modelled by a Gaussian disk with radius of 0.4$\deg$. 
The spectral of this component can be described by a broken power-law function with an index of $2.36 \pm 0.01$. The spectrum shape of region A reveals a significant pion-bump feature, which indicates the \gray\ emissions are from the interaction of hadronic CRs with ambient gas.  However, compared with the \gray\ emissions from other YMCs, the spectrum is significantly softer but is similar to the spectral shape of the Galactic diffuse emission  and the emissions from \etacarina\ itself. So it is possible the derived emission in this region are still significantly contaminated by the \etacarina\ which is by far brighter. Another possibility is that the CO gas and \ion{H}{II} gas in this region are illuminated by the soft CR component accelerated by the \etacarina. In the latter, the derived $W_{\rm p}$ is $10^{48}~\rm erg$ for the protons above 2 GeV, considering wind power of the \etacarina\ of $10^{38} ~\rm erg/s$ and a acceleration efficiency of 10\%. The CR injection power can be $P_{\rm CR} \sim 10^{37} ~\rm erg/s$. Taking into account the size of region A of about $l=15 ~\rm pc$  ($0.4\deg$ in $2.3~\rm kpc  $), the required diffusion coefficient in this region can be estimated as $D \sim \frac{l}{4T} $, where the confinement time $T$ can be estimated as  $W_{\rm p}/P_{\rm CR} \sim 10^{11} \rm s$. The derived $D$ is $5\times 10^{27} ~\rm cm/s$, which is one order of magnitude smaller than the average value in the  galactic plane. Such a suppression of diffusion coefficient is also found in other regions near CR source \citep{Aharonian19}. We note that other than \etacarina\ other massive stars in the young massive cluster Tr14 and Tr 16 may also accelerate CRs, which will loose our constraints on the diffusion coefficient. 
%Above 30 GeV there is no emission, and the upper limit reveals a spectral cutoff or break. The \gray\ spectrum can be described by leptonic and hadronic models alike. However, we argue that the most probable origin is the interaction of the accelerated protons in the YMCs with ambient gas. Since the central point source is very significant, we cannot rule out that region A may be contributed by \etacarina. The derived total CR proton energy is estimated to be as high as $\sim 3.3 \times\ 10^{48}\ \rm erg$.

The GeV \gray\ emission from the northern part of the CNC (region B) can be modelled as a 0.75$\deg$ Gaussian disk. 
The \gray\ emission has a hard spectrum with a spectral index of about $2.12 \pm 0.02$. Although the spectrum of region B can be fit by both leptonic and hadronic scenarios, the high gas density and good spatial correlation between \gray\ emission and molecular gas strongly favor a hadronic origin. In this case the hard \gray\ spectrum and the hard spectrum of parent CRs are similar to those in other YMC systems. A natural acceleration site of the CRs are the YMC Tr 14 and Tr 16, in this case we plot the radial distribution of CRs in region A and region B, which we found are not consistent with the $1/r$ distributions measured near other YMCs \citep{Aharonian19}. One possible explanation may be that the CR content in this system are indeed from some recent impulsive process rather than continuous injections. In this case the harder spectrum in region B compared with region A can also be explained as energy dependent propagation. Another possibility is that it is the young star cluster NGC 3324 or other unknown CR sources , rather than Tr14 and Tr 16, that accelerate the CRs in region B. Then the different spectrum in region A and region B can be attributed to the different CR injection spectrum in two systems. Indeed, in \citet{Yang18} the diffuse emission in this region are attributed to the CRs accelerated by the YMC Westerlund 2, which locates about $5~\rm kpc $ from the solar system. We note that after the detailed analysis in this paper we still cannot formally rule out such a possibility, although the projection distance of region B are significantly nearer with respect to \etacarina\ and Tr14/Tr16. The gas distribution derived near Westerlund 2 also reveal a peak in coincidence with region B \citep{Yang18}. Indeed the speed interval used in CNC ( $v_{\rm LSR}=[-32,-5]$ $\rm km\ s^{-1}$) and Westerlund 2 ( $v_{\rm LSR}=[-11,20]$ $\rm km\ s^{-1}$) are overlapped with each other, although their distance to the solar system are significantly different. More detailed studies on the gas distributions in this region may be required to pin down the origin of the diffuse \gray\ emissions in this region. 
%the CRs that are accelerated by the interaction of the young massive star clusters with the ambient gas. 
%

\section{Acknowledgements}
This work is supported by the National Natural Science Foundation of China (Grant No.12133003, 12103011, and U1731239), Guangxi Science Foundation (grant No. AD21220075). Rui-zhi Yang is supported by the NSFC under grants 11421303, 12041305 and the national youth thousand talents program in China.

\section{Data availability}
The \fermi\ data used in this work is publicly available, which is provided online by the NASA-GSFC Fermi Science Support Center\footnote{\url{ https://fermi.gsfc.nasa.gov/ssc/data/access/lat/}}.
We make use of the CO data\footnote{\url{ https://lambda.gsfc.nasa.gov/product/}} to derive the H$_{2}$. The data from \planck\ legacy archive\footnote{\url{ http://pla.esac.esa.int/pla/\#home}} are used to derive the \ion{H}{ii}.
The \ion{H}{i} data are from the HI4PI\footnote{\url{http://cdsarc.u-strasbg.fr/viz-bin/qcat?J/A+A/594/A116}}.

\bibliographystyle{mnras}
%\bibstyle{aa}
%\bibliographystyle{plain}
\bibliography{ms}

% Don't change these lines
\bsp	% typesetting comment
\label{lastpage}
\end{document}